\documentclass[10pt,conference]{IEEEtran}
\usepackage{epsfig,rotating,setspace,latexsym,amsmath,epsf,amssymb,amsfonts,bm,theorem,cite,authblk, bbm,color, multirow, caption, subcaption, booktabs}
\IEEEoverridecommandlockouts
\allowdisplaybreaks

\title{How to Discover Knowledge for FutureG: Contextual RAG and LLM Prompting for O-RAN} 

\author{Nathan Conger}
\author{Nathan Scollar}
\author{Kemal Davaslioglu}
\author{Yalin E. Sagduyu}
\author{Sastry Kompella}
\affil{\normalsize Nexcepta, Gaithersburg, MD, USA 
} 

\vspace{-2cm}

\date{}
\IEEEoverridecommandlockouts
\begin{document}
	
\maketitle
\vspace{-2cm}
\thispagestyle{empty}
	
\begin{abstract}
We present a retrieval-augmented question answering framework for 5G/6G networks, where the Open Radio Access Network (O-RAN) has become central to disaggregated, virtualized, and AI-driven wireless systems. While O-RAN enables multi-vendor interoperability and cloud-native deployments, its fast-changing specifications and interfaces pose major challenges for researchers and practitioners. Manual navigation of these complex documents is labor-intensive and error-prone, slowing system design, integration, and deployment. To address this challenge, we adopt Contextual Retrieval-Augmented Generation (Contextual RAG),
a strategy in which candidate answer choices guide document retrieval and chunk-specific context to improve large language model (LLM) performance. This improvement over traditional RAG achieves more targeted and context-aware retrieval, which improves the relevance of documents passed to the LLM, particularly when the query alone lacks sufficient context for accurate grounding. Our framework is designed for dynamic domains where data evolves rapidly and models must be continuously updated or redeployed, all without requiring LLM fine-tuning. We evaluate this framework using the ORAN-Benchmark-13K dataset, and  compare three LLMs, namely, Llama3.2, Qwen2.5-7B, and Qwen3.0-4B, across both Direct Question Answering (Direct Q\&A) and Chain-of-Thought (CoT) prompting strategies. We show that Contextual RAG consistently improves accuracy over standard RAG and base prompting, while maintaining competitive runtime and CO\textsubscript{2} emissions. These results highlight the potential of Contextual RAG to serve as a scalable and effective solution for domain-specific Q\&A in O-RAN and broader 5G/6G environments, enabling more accurate interpretation of evolving standards while preserving efficiency and sustainability.

\end{abstract}
	
	
\section{Introduction}
A question and answering (Q\&A) framework is increasingly necessary for 5G and future 6G systems, where the Open Radio Access Network (O-RAN) has become the foundation for disaggregated, virtualized, and AI-driven architectures. By introducing open interfaces and cloud-native principles, O-RAN enables multi-vendor interoperability and flexible deployments. However, its specifications evolve rapidly, producing large and technically dense standards that must align with 3GPP releases and define detailed requirements across functional splits and interfaces. Navigating these specifications presents significant challenges. Engineers must extract relevant information from large volumes of documents, reconcile frequent updates, and manage interdependencies across interfaces such as E2, A1, and O1. Manual approaches are slow, error-prone, and difficult to scale, often hindering reproducibility and delaying the deployment of interoperable solutions.

These challenges are expected to intensify as the wireless community transitions toward 6G, where O-RAN is expected to support even more heterogeneous, AI-native, and service-centric architectures. This motivates the need for intelligent Q\&A frameworks capable of retrieving, contextualizing, and reasoning over evolving standards at scale, reducing barriers to innovation, improving compliance, and accelerating both research and deployment. There is an increasing trend in curating high-quality, domain-specific datasets and integrating retrieval-based adaptation to bridge the gap between general-purpose LLM pre-training and the domain-specific requirements of technical documents for O-RAN, 3GPP, and other related 5G/6G ecosystems \cite{telcorag, teleqna, tspecllmopensourcedatasetllm, oran_gajjar2025oransight20foundationalllmsoran, gajjar2025oran}. 

Large Language Models (LLMs) such as Llama~\cite{llama3_2024}, Qwen~\cite{qwen3_2025}, and GPT have shown strong performance in tasks like Q\&A, reasoning, and code generation. However, their deployment in real-world settings is limited by outdated training corpora, inability to incorporate proprietary data, and difficulty correlating information across evolving sources. Retrieval-Augmented Generation (RAG)~\cite{rag_lewis2020retrieval} addresses these issues by retrieving relevant documents at query time and grounding outputs in up-to-date, verifiable information, thereby improving factual accuracy, reducing hallucinations, and avoiding costly retraining on sensitive data.

Recent studies have adapted RAG pipelines to specialized technical domains where documents are large, complex, and rapidly evolving. In telecommunications, Telco-RAG~\cite{telcorag} introduced domain-specific optimizations, such as glossary-based query augmentation, neural network routing, and hyperparameter tuning, to improve retrieval and reasoning over 3GPP standards. TSpec-LLM~\cite{tspecllmopensourcedatasetllm} complemented this effort by compiling an open-source dataset spanning all 3GPP releases from 1999–2023, preserving tables and formulas to enhance LLM understanding and downstream RAG tasks. In parallel, benchmark datasets have been developed to measure LLM competence in telecom-specific tasks: TeleQnA~\cite{teleqna} provided the first large-scale evaluation set across standards and research publications. ORAN-Bench-13K dataset \cite{gajjar2025oran} represents another milestone contribution to the O-RAN domain by providing a large-scale, domain-specific benchmark for evaluating LLMs on O-RAN specifications and research literature. ORANSight-2.0~\cite{oran_gajjar2025oransight20foundationalllmsoran} builds on this foundation by combining a RAG-based instruction-tuning pipeline (RANSTRUCT) with fine-tuned foundational models, while also introducing srsRANBench \cite{oran_gajjar2025oransight20foundationalllmsoran}, a dataset derived from the open-source srsRAN project, to support both text and code understanding in telecom tasks.


RAG performance depends heavily on retrieval strategy. In vanilla RAG, documents are retrieved solely from the user’s query, which can fail when queries lack specificity or when key context spans multiple documents, e.g., with time-sensitive references like ``the previous quarter." To address this, we adopt Contextual RAG~\cite{anthropic2024contextual}, which augments queries with chunk-level context and candidate answer choices. Originally applied to open-domain Q\&A, this approach enables more targeted and disambiguated retrieval without model fine-tuning, making it well-suited for O-RAN's rapidly evolving specifications.

In addition to retrieval strategies, prompting techniques play a critical role in the LLM performance. Direct Q\&A prompting aims to elicit concise answers based on the provided context, while Chain-of-Thought (CoT) prompting \cite{cot_wei2022chain} encourages step-by-step reasoning before arriving at a final answer. CoT has been shown to improve performance on reasoning tasks and reduce the retrieval failure rate \cite{anthropic2024contextual}.

To evaluate the interplay between retrieval and prompting strategies, we adopt multiple-choice Q\&A as a structured and measurable benchmark. This setup provides clear correctness criteria and reflects a common pattern in domain-specific applications such as regulatory compliance, technical documentation, and standards interpretation. We use the ORAN-Benchmark-13K dataset 
\cite{gajjar2025oran, oran_gajjar2025oransight20foundationalllmsoran} to compare retrieval methods (e.g., No RAG, Vanilla RAG, and Contextual RAG) and prompting techniques (Q\&A and CoT) across multiple state-of-the-art LLMs (Llama3.2, Qwen2.5-7B, and Qwen3.0-4B). Our results demonstrate that the Contextual RAG consistently outperforms Vanilla RAG and standard Q\&A prompting in accuracy while maintaining competitive runtime and CO\textsubscript{2} emissions, which demonstrates its value as a scalable solution for domain-specific Q\&A in O-RAN. While our evaluations focus on O-RAN, the proposed framework is generalizable and applicable to other technical, compliance-driven, and high-velocity information environments such as 3GPP.

The remainder of the paper is organized as follows. Sec.~\ref{sec:system_model} presents the Contextual RAG framework. Sec.~\ref{sec:oran_usecase} describes the O-RAN use case, and Sec.~\ref{sec:performance} reports the performance results. Sec.~\ref{sec:conclusion} concludes the paper.

\section{Contextual RAG Framework} \label{sec:system_model}

Our system model builds on the RAG framework to improve the LLM performance on domain-specific Q\&A. The main idea is to augment the LLM input with relevant textual context retrieved from a document corpus. We evaluate three configurations: no RAG (only internal knowledge of the LLM), standard RAG, and Contextual RAG.

\subsection{Retrieval-Augmented Generation (RAG)}

In the RAG setup, the user query is first embedded using a sentence-level embedding model and compared against a precomputed vector store of document chunks using cosine similarity. The top-$k$ most relevant chunks are retrieved and prepended to the prompt passed to the LLM. This allows the model to answer based on grounded, domain-specific evidence instead of relying solely on its pretraining.

\paragraph{Vector Store Construction} The vector store is built from 116 O-RAN specification documents. To prepare these documents for retrieval, they are split into semantically meaningful chunks using a recursive text splitting strategy. The splitter attempts to preserve semantic integrity by prioritizing paragraph boundaries, then sentences, and finally words to avoid hard splits that may disrupt meaning. 

\paragraph{Chunk Embeddings} Each chunk is embedded into a high-dimensional space using a transformer-based embedding model. 
While recursive splitting performs well on text-heavy documents, it may underperform on documents with multimodal elements such as tables, code blocks, or diagrams. More recent RAG frameworks typically include multimodal-aware chunkers or agent-based parsing to better preserve such structures. In terms of similarity measure, cosine similarity is used during the retrieval step to calculate the relevance between the query and document chunks.

\subsection{Bringing Context to RAG}

\begin{figure}[t!]
    \centering
    \includegraphics[width=0.9\linewidth]{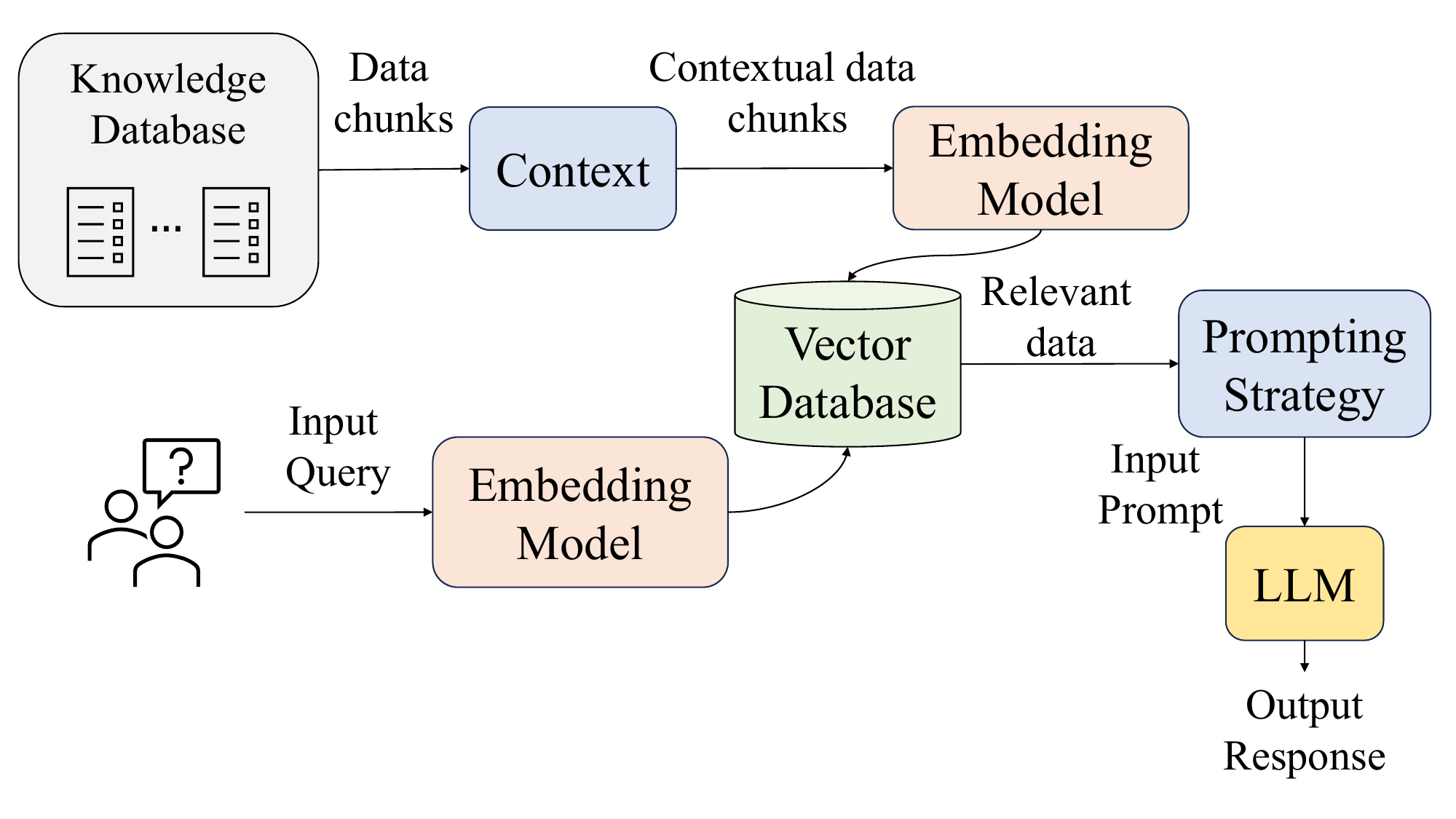}
    \caption{Contextual RAG retrieval framework.}
    \label{fig:corag}
\end{figure}

Contextual RAG extends standard RAG by incorporating chunk-specific context that explains the chunk using the context of the overall document. Our framework is illustrated in Fig.~\ref{fig:corag}, where we incorporate multiple-choice answer options into the retrieval query. Instead of retrieving documents using only the input query from the user, we form \emph{contextualized queries} that include each answer choice, which allows the retriever to focus on parts of the corpus that best distinguish among the options. This provides more targeted and context-aware retrieval, especially in cases where the question alone lacks enough specificity to ground the answer. As a result, the LLM receives input that is not only relevant to the question, but also discriminative with respect to potential answers.

For each multiple-choice question, we construct four queries by appending each candidate answer to the original question and pass them through the retriever. The top-$k$ chunks from all four queries are pooled, de-duplicated, and concatenated into the prompt passed to the LLM. This enhances the ability of the model to infer the correct answer by anchoring reasoning in choice-aware evidence.

\subsection{Prompting Strategies}

We consider two prompting methods for querying the LLM. The first, Direct Q\&A, uses a concise instruction prompt that asks the LLM to provide an answer based solely on the retrieved context. For example: ``Based on the following context, answer the question. Context: [retrieved text] Question: [question] Answer:” This format keeps the focus on selecting the correct choice without additional reasoning steps. The second, CoT, adopts a reasoning-oriented approach, which guides the LLM to explain its  steps before selecting an answer and closely mimics human logical deduction. For example: ``Based on the following context, think step-by-step to determine the answer. Context: [retrieved text] Question: [Question].” This format prompts the model to articulate its reasoning process before committing to a final answer.

\subsection{Evaluated LLMs}

We evaluate our framework using three open-weight LLMs that vary in architecture, size, and training methodology:

\begin{itemize}	
\item \textbf{Llama 3.2 3B} \cite{llama3_2024} is developed by Meta. This transformer-based model supports 128K context length, uses Grouped Query Attention (GQA), shared embeddings, and has a knowledge cutoff of December 2023.
	
\item \textbf{Qwen 2.5 7B} \cite{qwen2024} model includes a dense decoder-only transformer model with 28 layers, Rotary Positional Embeddings (RoPE), and has a 128K context length. Qwen~2.5 model is trained on 18 trillion tokens in 29 languages and has a knowledge cutoff of September 2024.
	
\item \textbf{Qwen~3.0 4B} \cite{qwen3_2025} is an enhanced version of Qwen~2.5. It was trained on 36 trillion tokens across 119 languages. Qwen~3.0 model introduces a \textit{thinking mode} for complex reasoning tasks and a \textit{thinking budget} mechanism that allows users to balance reasoning depth and inference efficiency. It is ideal for CoT prompting. While the exact knowledge cutoff date has not been disclosed, it is sometime in 2024.
	
\end{itemize}

\begin{table}[t!]
	\centering
        \footnotesize
	\caption{LLM performance across different benchmark tasks. Accuracy is represented in percentage.}
	\label{tab:model_performance}
	\begin{tabular}{lccc}
		\toprule
		\textbf{Task} & \textbf{Llama3.2 (3B)} & \textbf{Qwen2.5 (7B)} & \textbf{Qwen3.0 (4B)} \\
		\midrule
		MMLU  & 63.40 & 74.16 & 72.99 \\
		MATH  & 48.00 & 49.80 & 54.10 \\
		MGSM  & 58.20 & 63.60 & 67.74 \\
		\bottomrule
	\end{tabular}	
\end{table}

All three models demonstrate strong performance on general tasks, especially those involving Science, Technology, Engineering, and Mathematics (STEM), as well as multilingual benchmarks. Table~\ref{tab:model_performance} compares the performance of these three LLMs across three representative benchmarks: Massive Multitask Language Understanding (MMLU) that covers a broad range of academic and professional subjects; MATH that is a dataset of high school competition-level math problems that test symbolic and multi-step reasoning; and Multilingual Grade School Math (MGSM) that evaluates the ability of the model to solve math word problems in multiple languages. These benchmarks collectively highlight the generalization and reasoning capabilities of the evaluated models. 

\section{O-RAN Use Case} \label{sec:oran_usecase}

As a representative domain-specific use case, we evaluate our framework using the ORAN-Bench-13K dataset \cite{oran_gajjar2025oransight20foundationalllmsoran}. This dataset comprises 13,952 multiple-choice questions derived from 116~O-RAN specification documents. It also includes a vector store for retrieval that is constructed from the same documents and preprocessed using a recursive semantic chunking strategy. For a fair comparison, we use the same vector store provided by the dataset. The ORAN-Bench-13K dataset was generated through a multi-step pipeline involving both generation and validation by large language models:
\begin{itemize}
	\item \textbf{Chunking}: The O-RAN documents were initially split into overlapping chunks of size 1536~characters with a 256-character overlap.	
	\item \textbf{Question Generation}: A Generator LLM produced a multiple-choice question (with four answer options and a labeled correct answer) based on each chunk.	
	\item \textbf{Answer Validation}: A separate Validator LLM independently answered the same question using the chunk as context. If both models agreed on the correct answer, the question was accepted; otherwise, it was discarded.
	
	\item \textbf{Difficulty Categorization}: Accepted questions were classified by a third Categorizer LLM into one of three difficulty levels, namely, easy, immediate, and difficult.
\end{itemize}

The final dataset includes 1,139 easy, 9,570 medium, and 3,243 difficult questions. The diversity and rigor of this benchmark make it an ideal testbed for studying retrieval and prompting strategies in high-precision, technical domains.

While ORAN-Bench-13K serves as the primary evaluation setting in this work, the proposed framework is applicable to other evolving, document-heavy domains such as cybersecurity, legal compliance, and standards-driven engineering.

Our implementation uses the Ollama library for fast LLM inference and the LangChain library to manage the retrieval pipeline, construct prompts, and coordinate interactions between the retriever and the LLM. For vector storage, the ORAN-Bench-13K dataset is indexed using the FAISS library and we use the same retriever for a fair evaluation.

\section{Performance Evaluation \label{sec:performance}}
To evaluate the performance of Contextual RAG, we assess the multiple-choice accuracy, inference latency, and CO\textsubscript{2} emissions which are described next.
\begin{table}[t!]	
	\centering
	\caption{Multi-choice Q\&A accuracy performance of various LLMs across retrieval and prompting strategies.}
    \label{table:accuracy_overall}
	\resizebox{\columnwidth}{!}{
		\begin{tabular}{lllcclll}
			\toprule
			\textbf{LLM Model} & \textbf{Retrieval} & \textbf{Prompt} & \textbf{Easy} & \textbf{Medium} & \textbf{Hard} \\ 
			\midrule
			ORANSight-Qwen-2.5:7B & RAG & Q\&A & 0.788 & 0.720 & 0.696 \\
			ChatGPT o4-mini & No RAG & Q\&A & 0.766 & 0.727 & 0.677 \\
			ChatGPT o4 & No RAG & Q\&A & \textbf{0.792} & \textbf{0.760} & \textbf{0.693} \\
			Gemini 1.5:8B & No RAG & Q\&A & 0.723 & 0.665 & 0.631 \\
			Gemini 1.5 & No RAG & Q\&A & 0.743 & 0.707 & 0.669 \\	
			\midrule	
			\multirow{6}{*}{Llama3.2}
			& No RAG & Q\&A & 0.6883 & 0.6241 & 0.5596 \\
			& No RAG & CoT & 0.6177 & 0.5760 & 0.5115 \\
			& RAG & Q\&A & 0.6932 & 0.6564 & 0.6212 \\
			& RAG & CoT & 0.7153 & 0.6934 & 0.6987 \\
			& Co-RAG & Q\&A & 0.7171 & 0.6700 & 0.6315 \\
			& Co-RAG & CoT & \textbf{0.7528} & \textbf{0.7293} & \textbf{0.7139} \\
			\midrule
			\multirow{6}{*}{Qwen 2.5:7B} 
			& No RAG & Q\&A & 0.7480 & 0.7044 & 0.6390 \\
			& No RAG & CoT & 0.7223 & 0.6582 & 0.5819 \\
			& RAG & Q\&A & 0.7813 & 0.7708 & 0.7423 \\
			& RAG & CoT & 0.7981 & 0.7711 & 0.7502 \\
			& Co-RAG & Q\&A & 0.8206 & 0.7863 & 0.7637 \\
			& Co-RAG & CoT & \textbf{0.8666} & \textbf{0.8243} & \textbf{0.7928} \\
			\midrule
			\multirow{6}{*}{Qwen 3.0:4B}
			& No RAG & Q\&A & 0.7524 & 0.6949 & 0.6313 \\
			& No RAG & CoT & 0.7436 & 0.6970 & 0.6257 \\
			& RAG & Q\&A & 0.8314 & 0.8189 & 0.7962 \\
			& RAG & CoT & 0.8393 & 0.8162 & 0.8003 \\
			& Co-RAG & Q\&A & \textbf{0.8824} & \textbf{0.8718} & 0.8261 \\
			& Co-RAG & CoT & 0.8761 & 0.8687 & \textbf{0.8475} \\
			\bottomrule
		\end{tabular}
	}	
\end{table}


\subsection{Accuracy}
Table~\ref{table:accuracy_overall} presents the multiple-choice Q\&A accuracy of various LLMs across retrieval frameworks and prompting strategies for each difficulty level. In this table, we use the notation Co-RAG to denote Contextual RAG. For benchmarking, we also include the results reported in~\cite{oran_gajjar2025oransight20foundationalllmsoran}, which provide strong reference points from models such as ORANSight-Qwen-2.5~7B, ChatGPT o4-mini, ChatGPT~o4, Gemini~1.5~8B, and Gemini~1.5. Among these benchmark models, ChatGPT~o4 achieves the highest accuracy with 0.792, 0.760, and 0.693 for increasing difficulty levels, respectively.

When comparing Llama~3.2, Qwen~2.5~7B, and Qwen~3.0~4B, several trends emerge. First, when relying solely on the internal knowledge of the LLM (no retrieval), the CoT prompting strategy generally does not improve performance, and in many cases, it reduces accuracy by up to 7\%. As an example, Qwen~2.5 with direct Q\&A achieves 0.6883 in Easy questions, but reduces to 0.6177 with CoT without RAG. However, for the frameworks that have retrieval-augmented strategies (RAG and Contextual RAG), CoT consistently improves multiple choice accuracy over Direct Q\&A for Llama~3.2 and Qwen~2.5~7B. For instance, Llama~3.2 RAG improved from 0.6932, 0.6564, and 0.6212 to 0.7153, 0.6934, and 0.6987, respectively, across difficulty levels. For these models, Contextual RAG consistently outperforms standard RAG. For instance, Qwen~2.5 with Co-RAG and CoT prompting achieves accuracies of 0.8666, 0.8243, and 0.7928 on the easy, medium, and hard sets, respectively, compared to 0.8206, 0.7863, and 0.7637 with direct Q\&A prompting.

For Qwen~3.0~4B, prompting strategy has little effect since this model already incorporates CoT reasoning in its built-in ``thinking” mode. In contrast, the retrieval framework plays a more decisive role: Contextual RAG consistently outperforms standard RAG across all difficulty levels. For example, with CoT prompting, RAG achieves accuracies of 0.8393, 0.8162, and 0.8003 on the easy, medium, and hard sets, respectively, whereas Contextual RAG attains 0.8761, 0.8687, and 0.8475. 

Overall, the Qwen~3.0~4B model with Contextual RAG under both direct Q\&A and CoT prompting strategies outperforms all comparable benchmark models reported in~\cite{oran_gajjar2025oransight20foundationalllmsoran} by an absolute margin of 8–15\%. For instance, while ChatGPT~o4 achieves accuracies of 0.792, 0.760, and 0.693 on the easy, medium, and hard sets, respectively, Contextual RAG with direct Q\&A attains 0.8824, 0.8718, and 0.8261.

\begin{table}[t!]
	\centering        
	\caption{Latency performance of various LLMs across retrieval and prompting strategies.}
    \label{table:latency_overall}
	\footnotesize
	\resizebox{\columnwidth}{!}{%
		\begin{tabular}{lllcccccc}
			\toprule
			\textbf{LLM Model} & \textbf{Retrieval} & \textbf{Prompt} & 
			\multicolumn{2}{c}{\textbf{Easy}} & 
			\multicolumn{2}{c}{\textbf{Medium}} & 
			\multicolumn{2}{c}{\textbf{Hard}} \\
			\cmidrule(r){4-5} \cmidrule(r){6-7} \cmidrule(r){8-9}
			& & & Mean & SD & Mean & SD & Mean & SD \\
			\midrule
			
			\multirow{6}{*}{Llama3.2}
			& No RAG & Q\&A & 0.34 & 0.63 & 0.42 & 0.90 & 0.62 & 1.74 \\
			& No RAG & CoT & 5.85 & 1.45 & 6.44 & 1.66 & 6.79 & 2.11 \\
			& RAG & Q\&A & 0.96 & 0.42 & 0.98 & 0.31 & 1.10 & 0.63 \\
			& RAG & CoT & 6.21 & 1.86 & 6.47 & 2.20 & 7.10 & 2.64 \\
			& Co-RAG & Q\&A & 0.89 & 0.24 & 0.94 & 0.28 & 1.02 & 0.42 \\
			& Co-RAG & CoT & 5.89 & 1.98 & 6.11 & 2.43 & 6.49 & 2.57 \\
			\midrule
			
			\multirow{6}{*}{Qwen 2.5:7B}
			& No RAG & Q\&A & 0.29 & 0.03 & 0.31 & 0.03 & 0.32 & 0.04 \\
			& No RAG & CoT & 10.74 & 2.18 & 11.33 & 2.30 & 12.23 & 2.92 \\
			& RAG & Q\&A & 1.64 & 0.41 & 1.79 & 0.47 & 1.98 & 0.47 \\
			& RAG & CoT & 11.71 & 2.88 & 12.40 & 2.80 & 13.16 & 3.60 \\
			& Co-RAG & Q\&A & 1.78 & 0.36 & 1.81 & 0.36 & 2.02 & 0.44 \\
			& Co-RAG & CoT & 10.21 & 2.51 & 10.62 & 2.70 & 11.78 & 4.15 \\
			\midrule
			
			\multirow{6}{*}{Qwen 3.0:4B}
			& No RAG & Q\&A & 20.54 & 15.14 & 26.12 & 18.87 & 31.45 & 25.44 \\
			& No RAG & CoT & 21.39 & 15.27 & 27.20 & 19.22 & 31.82 & 23.56 \\
			& RAG & Q\&A & 26.70 & 26.62 & 27.75 & 29.06 & 32.64 & 36.10 \\
			& RAG & CoT & 31.39 & 30.53 & 32.49 & 35.05 & 32.33 & 36.08 \\
			& Co-RAG & Q\&A & 23.88 & 26.66 & 26.45 & 28.65 & 31.90 & 37.57 \\
			& Co-RAG & CoT & 27.84 & 29.56 & 31.91 & 35.87 & 32.67 & 37.10 \\
			\bottomrule
		\end{tabular}%
	}
\end{table}

\begin{figure}[t!]
    \centering
    \includegraphics[width=0.7\linewidth]{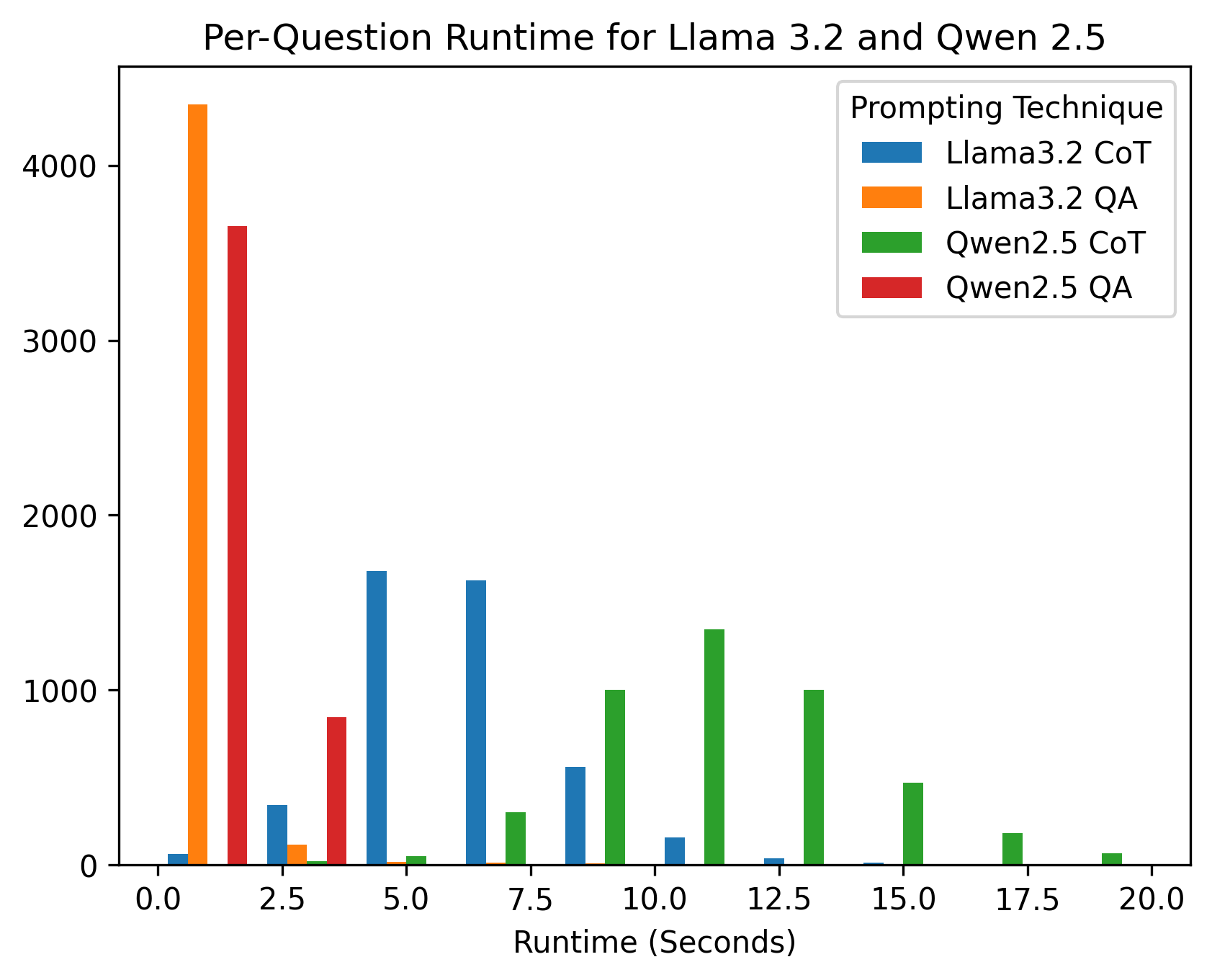}
    \caption{Inference run times across different prompting strategies for the Llama~3.2 and Qwen~2.5 models.}
    \label{fig:pdf_runtime_llama_qwen}
\end{figure}

\begin{figure}[t!]
    \centering
    \includegraphics[width=0.7\linewidth]{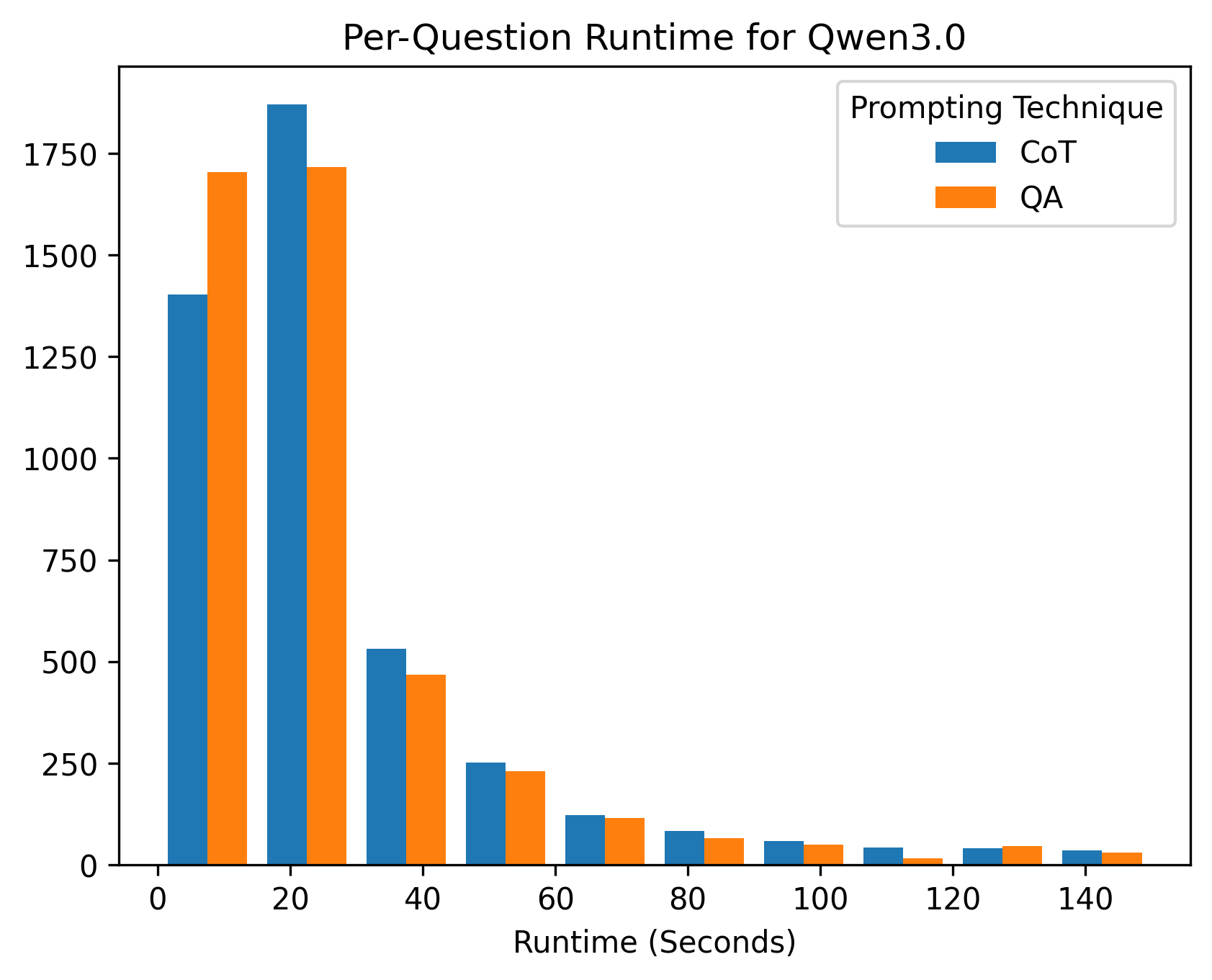}
    \caption{Inference run times across different prompting strategies for the Qwen~3.0 Model.}
    \label{fig:pdf_runtime_qwen3}
\end{figure}

\subsection{Latency}
Inference-time latency is a key performance metric in Q\&A systems. In our evaluations, we measured only the LLM inference time, excluding document retrieval. This exclusion is justified because the retrieval pipeline is identical across all models, retrieval methods, and prompting strategies, and therefore does not influence comparative latency results. For each difficulty level, a fixed set of 500 questions was used across all experiments to ensure fairness and reproducibility.

Table~\ref{table:latency_overall} summarizes the mean and standard deviation (SD) of inference latencies for all three LLMs under different retrieval and prompting strategies. Figs.~\ref{fig:pdf_runtime_llama_qwen} and \ref{fig:pdf_runtime_qwen3} present the probability distribution of these models for different prompting strategies. The largest latency difference appears in Llama~3.2 and Qwen~2.5, where CoT prompting is 5–10$\times$ slower than direct Q\&A. This slowdown is expected, as CoT responses involve step-by-step reasoning and produce substantially more tokens. In contrast, Qwen~3.0 exhibits only a marginal CoT overhead which is about 1.08$\times$ longer than direct Q\&A, which suggests that its built-in \emph{thinking} mode mitigates much of the additional computational cost.

\begin{figure}[tb!]
    \centering
    \includegraphics[width=0.7\linewidth]{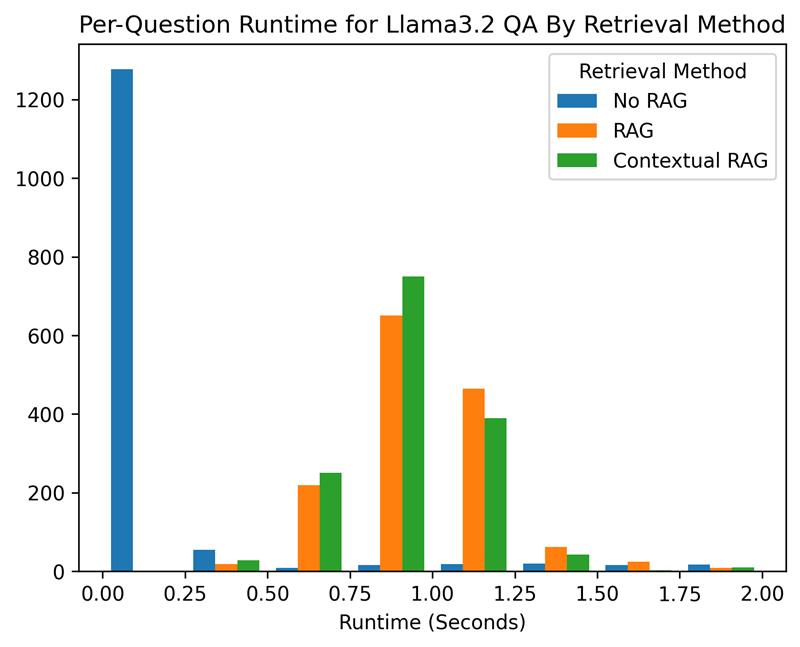}
    \caption{Inference run times across retrieval methods for the Llama~3.2 model.}
    \label{fig:runtime_llama32_per_retrieval_method}
\end{figure}

\begin{figure}[tb!]
    \centering
    \includegraphics[width=0.7\linewidth]{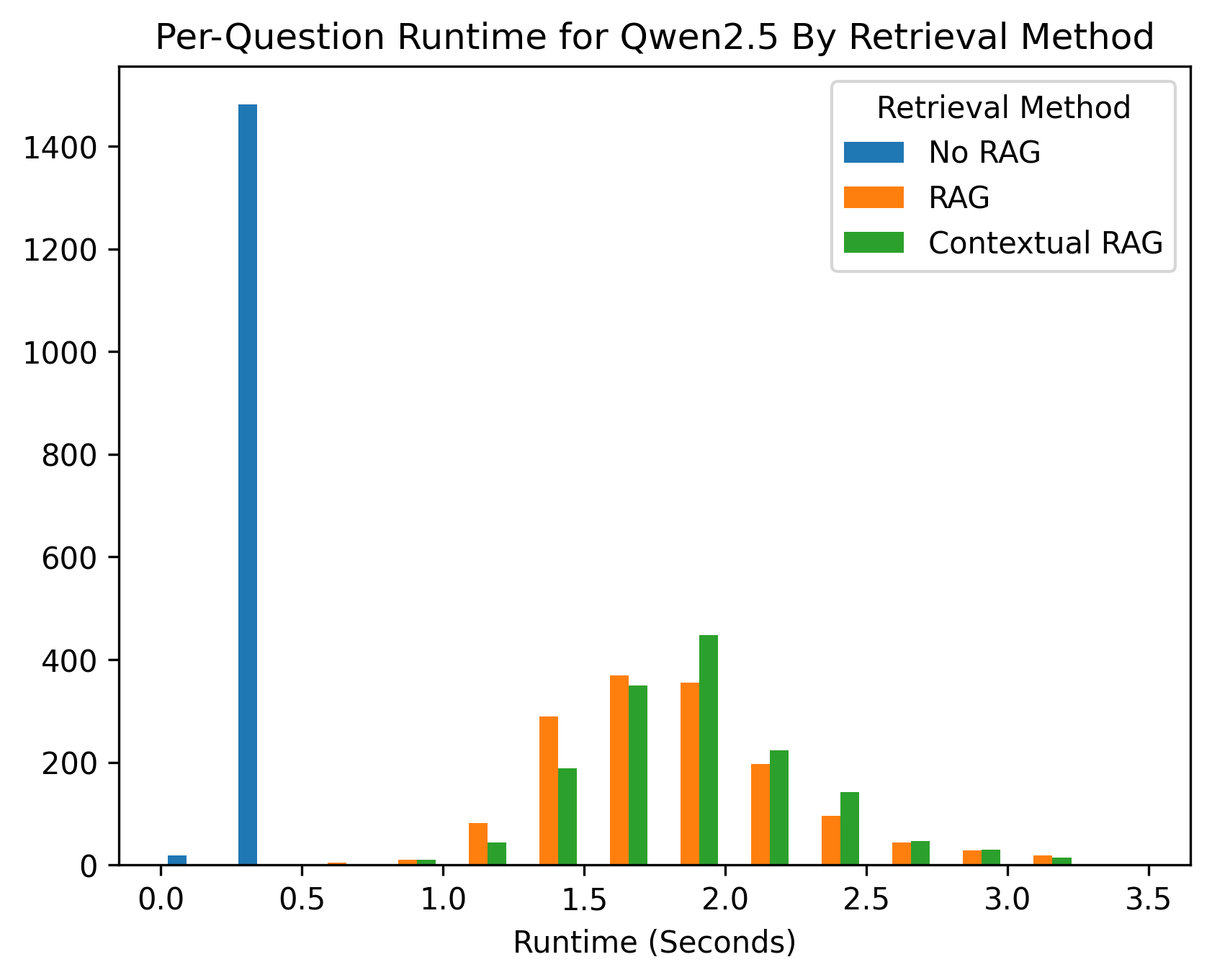}
    \caption{Inference run times across retrieval methods for the Qwen~2.5 model.}
    \label{fig:runtime_qwen25_per_retrieval_method}
\end{figure}

\begin{figure}[tb!]
    \centering
    \includegraphics[width=0.7\linewidth]{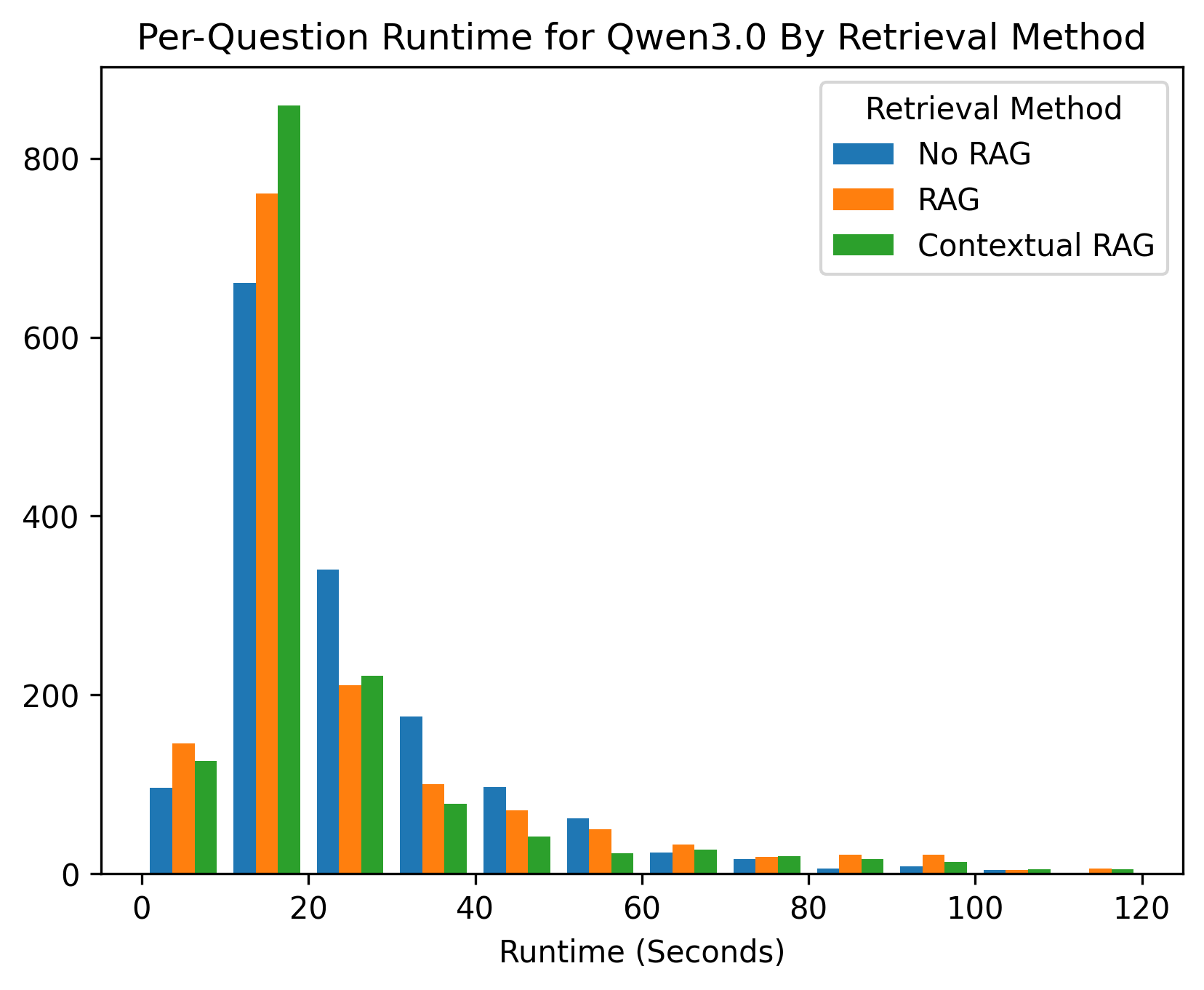}
    \caption{Inference run times across retrieval methods for the Qwen~3.0 model.}    \label{fig:runtime_qwen3_per_retrieval_method}
\end{figure}

Figs.~\ref{fig:runtime_llama32_per_retrieval_method}-\ref{fig:runtime_qwen3_per_retrieval_method} present the histogram of Llama~3.2, Qwen~2.5, and Qwen~3.0 models, respectively, for No RAG (only internal knowledge), RAG, and Contextual RAG frameworks. When these figures are studied with Table~\ref{table:latency_overall}, we observe that retrieval method also affects latency, though its impact varies based on the prompting style. With Q\&A prompting, Llama~3.2 runtime increases from 0.46 s without RAG to 1.02 s with RAG and 0.95 s with Contextual RAG—slowdowns of 2.21$\times$ and 2.06$\times$, respectively. Qwen~2.5 shows an even stronger effect, where the latency is 0.31~sec with No RAG versus 1.83~sec with RAG and 1.87~sec with Contextual RAG, corresponding to slowdowns of 5$\times$ and 6$\times$, respectively. This creates the bimodal runtime distribution depicted in Fig.~\ref{fig:pdf_runtime_llama_qwen}, where No RAG forms the lower-latency mode and the RAG variants form the higher-latency mode. In contrast, Qwen~3.0 with Q\&A prompting has a much smaller retrieval impact, with average runtimes of 26.04, 29.03, and 27.73~sec for No RAG, RAG, and Contextual RAG, respectively.

With CoT prompting, retrieval choice has a smaller impact for all models. For example, Llama~3.2 averages 6.36~sec with No RAG, 6.60~sec with RAG, and 6.16~sec with Contextual RAG, while Qwen~2.5 averages 11.44~sec, 12.43~sec, and 10.87~sec, respectively. Qwen~3.0 follows the same trend: 26.80~sec with No RAG, 32.07~sec with RAG, and 30.91~sec with Contextual RAG.

To test statistical significance, we applied an Analysis of Variance (ANOVA)~\cite{girden1992anova} test separately for Q\&A and CoT on each model. In all six cases, retrieval method had a statistically significant effect on latency ($p < 0.05$), with No RAG consistently fastest, and RAG and Contextual RAG were slower to varying degrees depending on the model.

While Qwen~3.0 exhibits higher inference latency, this overhead is expected for compact models with built-in reasoning mechanisms, and our results show that Contextual RAG still delivers accuracy gains without disproportionately increasing runtime. We also observe that latency correlates with question difficulty. Averaged across all models and prompting styles, easy questions took 11.56~sec, medium questions 12.92~sec, and hard questions 14.29~sec. An ANOVA test confirmed these differences are statistically significant ($p<0.01$). As shown in Table~\ref{table:runtime_breakdown}, easy questions dominate the lowest runtime thresholds, whereas hard questions are far more likely to exceed 25~sec with 36.6\% more likely to fall in the 25–50~sec range and 54.9\% more likely to exceed 50~sec compared to easy questions. This consistent pattern across all models indicates that question difficulty is a strong predictor of runtime, which is independent of retrieval method or prompting strategy.


\subsection{CO2 Emissions}
Given the close relationship between computational time and energy consumption, we next examine the CO\textsubscript{2} emissions associated with each model, retrieval strategy, and prompting technique. Emissions were measured using the \texttt{codecarbon} Python library, with results reported in Table~\ref{table:co2_overall} are on a per-question basis across the three difficulty levels. Since the document retrieval process was identical across all experimental configurations, only LLM inference emissions were recorded.

The results reveal strong alignment between CO\textsubscript{2} emissions and inference latency trends. For Llama~3.2 and Qwen~2.5~7B, the CoT prompting strategy consistently produced significantly higher emissions (often 5–10$\times$ greater) than direct Q\&A, which mirrors the large runtime differences observed in the latency analysis. Higher question difficulty was also associated with higher emissions on average. Across models, Llama~3.2 exhibited the lowest emissions, while Qwen~3.0~4B produced the highest. Retrieval method choice also influenced emissions: No RAG consistently yielded the lowest values, while RAG and Contextual RAG incurred higher emissions, with the maximum depending on the specific model and retrieval strategy.


Overall, our findings confirm that both model architecture and prompting strategy are major determinants of emissions, with CoT consistently increasing computational cost relative to Q\&A. The magnitude of this increase is model-dependent: Qwen~3.0~4B, which defaults to a “thinking mode” that implicitly incorporates CoT reasoning, exhibited minimal emission differences between Q\&A and CoT. This aligns with the similar accuracy and latency results of the model across the two prompting strategies, in contrast to the more substantial trade-offs observed in Llama~3.2 and Qwen~2.5~7B.




\begin{table}[t!]
	\centering
	\caption{Breakdown of question runtime percentages by difficulty and threshold. Data is aggregated over all LLM models and prompting techniques.}
    \label{table:runtime_breakdown}
	\footnotesize
	\begin{tabular}{lccc}
		\toprule
		\textbf{Runtime Threshold} & \textbf{Easy} & \textbf{Medium} & \textbf{Hard} \\
		\midrule
		Runtime $<$ 5~sec            & 38.04\% & 37.00\% & 36.74\% \\
		5~sec $<$ Runtime $\leq$ 10~sec  & 20.86\% & 18.99\% & 17.31\% \\
		10~sec $<$ Runtime $\leq$ 15 sec & 20.70\% & 19.71\% & 19.33\% \\
		15~sec $<$ Runtime $\leq$ 20~sec & 8.07\%  & 9.37\%  & 9.59\%  \\
		20~sec $<$ Runtime $\leq$ 25~sec & 3.42\%  & 3.90\%  & 4.27\%  \\
		25~sec $<$ Runtime $\leq$ 50~sec & 5.74\%  & 6.89\%  & 7.84\%  \\
		Runtime $>$ 50~sec           & 3.17\%  & 4.14\%  & 4.91\%  \\
		\bottomrule
	\end{tabular}	
\end{table}

\begin{table}[t!]
	\centering
	\caption{Per-question CO\textsubscript{2} emissions (grams) for different LLMs, retrieval strategies, and prompting techniques across dataset difficulty levels.}
    \label{table:co2_overall}
	\resizebox{\columnwidth}{!}{%
		\begin{tabular}{lllcccccc}
			\toprule
			\textbf{LLM Model} & \textbf{Retrieval} & \textbf{Prompt} & 
			\multicolumn{2}{c}{\textbf{Easy}} & 
			\multicolumn{2}{c}{\textbf{Medium}} & 
			\multicolumn{2}{c}{\textbf{Hard}} \\
			\cmidrule(r){4-5} \cmidrule(r){6-7} \cmidrule(r){8-9}
			& & & Mean & SD & Mean & SD & Mean & SD \\
			\midrule
			
			\multirow{6}{*}{Llama3.2} 
			& No RAG        & Q\&A  & 0.02 & 0.01 & 0.02 & 0.02 & 0.03 & 0.04 \\
			& No RAG        & CoT & 0.16 & 0.04 & 0.18 & 0.04 & 0.18 & 0.05 \\
			& RAG           & Q\&A  & 0.03 & 0.01 & 0.04 & 0.01 & 0.04 & 0.02 \\
			& RAG           & CoT & 0.17 & 0.05 & 0.18 & 0.06 & 0.19 & 0.07 \\
			& Co-RAG & Q\&A  & 0.03 & 0.01 & 0.03 & 0.01 & 0.04 & 0.01 \\
			& Co-RAG & CoT & 0.16 & 0.05 & 0.17 & 0.06 & 0.18 & 0.07 \\
			
			\midrule
			\multirow{6}{*}{QWEN 2.5 7B}
			& No RAG        & Q\&A  & 0.02 & $<$0.01 & 0.02 & $<$0.01 & 0.02 & $<$0.01 \\
			& No RAG        & CoT & 0.30 & 0.06    & 0.31 & 0.06    & 0.34 & 0.08 \\
			& RAG           & Q\&A  & 0.05 & 0.01    & 0.06 & 0.01    & 0.06 & 0.01 \\
			& RAG           & CoT & 0.31 & 0.08    & 0.33 & 0.07    & 0.35 & 0.09 \\
			& Co-RAG & Q\&A  & 0.06 & 0.01    & 0.06 & 0.01    & 0.06 & 0.01 \\
			& Co-RAG & CoT & 0.28 & 0.07    & 0.29 & 0.07    & 0.32 & 0.11 \\
			
			\midrule
			\multirow{6}{*}{QWEN 3.0 4B}
			& No RAG        & Q\&A  & 0.54 & 0.39 & 0.68 & 0.48 & 0.81 & 0.65 \\
			& No RAG        & CoT & 0.55 & 0.40 & 0.71 & 0.49 & 0.82 & 0.60 \\
			& RAG           & Q\&A  & 0.69 & 0.68 & 0.72 & 0.74 & 0.84 & 0.92 \\
			& RAG           & CoT & 0.82 & 0.78 & 0.84 & 0.89 & 0.84 & 0.92 \\
			& Co-RAG & Q\&A  & 0.62 & 0.68 & 0.71 & 0.91 & 0.83 & 0.97 \\
			& Co-RAG & CoT & 0.73 & 0.75 & 0.83 & 0.91 & 0.85 & 0.95 \\
			
			\bottomrule
		\end{tabular}%
	}	
\end{table}

\section{Conclusion} \label{sec:conclusion}
In this paper, we presented a retrieval-augmented Q\&A framework that employs Contextual RAG to improve the accuracy of LLMs without requiring fine-tuning. Through extensive evaluation on the ORAN-Bench-13K dataset across three open-weight LLMs, namely, Llama 3.2, Qwen 2.5-7B, and Qwen 3.0-4B, and showed that all retrieval-augmented methods outperformed No RAG technique that relies on internal knowledge of the LLM. Among the approaches we studied, Contextual RAG with Q\&A and CoT prompting strategies consistently provided the highest accuracy and outperformed the strong benchmarks listed in \cite{oran_gajjar2025oransight20foundationalllmsoran} by an absolute margin of 8-15\%. This points out that by enabling more targeted and relevant retrievals through the use of candidate answer choices during the retrieval stage, LLMs can better differentiate between the choices and improve their multiple choice decision making. We also found that inference time varied significantly depending on the model architecture, with Qwen~3.0-4B exhibiting the longest latency but also showing robustness across prompting strategies. Notably, Qwen 3.0's internal CoT mechanism resulted in relatively small performance differences between Direct Q\&A and CoT prompting. In contrast, for other models, the use of CoT prompting without retrieval often underperformed compared to Direct Q\&A combined with retrieval. These results highlight the importance of both retrieval and prompting strategies and suggest that Contextual RAG offers a scalable and effective solution for domain-specific Q\&A tasks in dynamic environments.


\bibliographystyle{IEEEtran}
\bibliography{references}
	
\end{document}